\documentclass[journal,twoside]{IEEEtran}

\ifCLASSINFOpdf
\usepackage[pdftex]{graphicx}

\else

\fi

\usepackage[cmex10]{amsmath}
\usepackage{amssymb}   
\usepackage{amsxtra}
\usepackage{amscd}
\usepackage{amsthm}
\usepackage{textcomp}
\usepackage{graphicx}  
\usepackage{color}

\usepackage{balance}

\usepackage{flushend}

\usepackage{array}  

\usepackage{multirow}  

\usepackage{flushend}
\usepackage{amsmath}
\usepackage{cite}

\usepackage{url}

\usepackage{color,soul} 
\soulregister\cite7
\soulregister\ref7
\soulregister\pageref7

\begin{document}
\title{Reconfigurable Intelligent Surfaces for 6G:\\Emerging Hardware Architectures, Applications,\\ and Open Challenges}

\author{Ertugrul Basar,~\IEEEmembership{Fellow,~IEEE,}
        George C. Alexandropoulos,~\IEEEmembership{Senior~Member,~IEEE,}\\
        Yuanwei Liu,~\IEEEmembership{Fellow,~IEEE,}
        Qingqing Wu,~\IEEEmembership{Senior~Member,~IEEE,}
        Shi Jin,~\IEEEmembership{Fellow,~IEEE,}\\
        Chau Yuen,~\IEEEmembership{Fellow,~IEEE,}
        Octavia A. Dobre,~\IEEEmembership{Fellow,~IEEE,}
        and~Robert Schober,~\IEEEmembership{Fellow,~IEEE}
\thanks{E. Basar is with the Communications Research and Innovation Laboratory (CoreLab), Department of Electrical and Electronics Engineering, Ko\c{c} University, Sariyer 34450, Istanbul, Turkey (e-mail: ebasar@ku.edu.tr).}
\thanks{G. Alexandropoulos is with the Department of Informatics and Telecommunications, National and Kapodistrian University of Athens Greece (e-mail: alexandg@di.uoa.gr).}
\thanks{Y. Liu is with the School of Electronic Engineering and Computer Science, Queen Mary University of London, London E1 4NS, UK (email: yuanwei.liu@qmul.ac.uk).}
\thanks{Q. Wu is with the Department of Electronic Engineering, Shanghai Jiao Tong University, 200240, China (email: qingqingwu@sjtu.edu.cn).}
\thanks{S. Jin is with the National Mobile Communications Research Laboratory, Southeast University, Nanjing 210096, China (e-mail: jinshi@seu.edu.cn).}
\thanks{C. Yuen is with the School of Electrical and Electronics Engineering, Nanyang Technological University, 50 Nanyang Ave, Singapore 639798 (email: chau.yuen@ntu.edu.sg).}
\thanks{O. A. Dobre is with Department of Electrical and Computer Engineering, Memorial University, 240 Prince Philip Dr., St. John’s, NL A1B 3X5 Canada (e-mail: odobre@mun.ca).}
\thanks{R. Schober is with the Institute for Digital Communications, FriedrichAlexander University of Erlangen-Nuremberg, Erlangen 91058, Germany (email: robert.schober@fau.de).}
}

\IEEEspecialpapernotice{(Invited Paper)}

\maketitle
\begin{abstract}
Reconfigurable intelligent surfaces (RISs) are rapidly gaining prominence in the realm of fifth generation (5G)-Advanced, and predominantly, sixth generation (6G) mobile networks, offering a revolutionary approach to optimizing wireless communications. This article delves into the intricate world of the RIS technology, exploring its diverse hardware architectures and the resulting versatile operating modes. These include RISs with signal reception and processing units, sensors, amplification units, transmissive capability, multiple stacked components, and dynamic metasurface antennas. Furthermore, we shed light on emerging RIS applications, such as index and reflection modulation, non-coherent modulation, next generation multiple access, integrated sensing and communications (ISAC), energy harvesting, as well as aerial and vehicular networks. These exciting applications are set to transform the way we will wirelessly connect in the upcoming era of 6G. Finally, we review recent experimental RIS setups and present various open problems of the overviewed RIS hardware architectures and their applications. From enhancing network coverage to enabling new communication paradigms, RIS-empowered connectivity is poised to play a pivotal role in shaping the future of wireless networking. This article unveils the underlying principles and potential impacts of RISs, focusing on cutting-edge developments of this physical-layer smart connectivity technology.\end{abstract}


\IEEEpeerreviewmaketitle

\newpage
\section{Introduction}
\IEEEPARstart{W}{ireless} connectivity empowered by reconfigurable intelligent surfaces (RISs) has emerged as a groundbreaking technology in the ever-evolving landscape of sixth generation (6G) mobile networks \cite{Basar_2019,huang2019reconfigurable}. In a world where seamless and efficient wireless communications is increasingly essential, RISs offer a compelling solution by deliberately manipulating the radio propagation environment. This article provides an in-depth exploration of the recent developments in this emerging technology, delving into its multifaceted dimensions, from its diverse architectural considerations to its different operating modes. As fifth generation (5G)-Advanced is being standardized and 6G networks loom on the horizon, the relevance and significance of RISs are becoming increasingly apparent.

One of the central aspects we examine, in this article, is the hardware structure of an RIS, overviewing the latest advances in the technology's core hardware components. These metasurfaces contain many tiny unit elements that can be controlled to manipulate their impinging electromagnetic (EM) waves. The intricate interplay of these elements opens up a world of possibilities for customizing wireless communications. In this context, understanding the underlying RIS hardware components is crucial to realizing the potential of the technology in 6G networks. Notably, we provide detailed coverage of emerging RIS architectures, including RISs with signal reception and processing units, sensors, amplification units, transmissive capability, multiple stacked components (i.e., 3D RIS-based structures), and dynamic metasurface antennas.

In addition to the RIS hardware architectures, the article illuminates the various operating modes resulting from them. These modes enable RISs to serve diverse purposes, from improving network coverage in challenging environments to supporting entirely new communications and sensing paradigms. By allowing real-time adaptability, the RIS technology empowers network operators to optimize the wireless environments where their systems operate dynamically~\cite{ris_5Gadvanced}. This adaptability is crucial in addressing the ever-increasing demands of 6G networks. At this point, we shift our focus into emerging reflection, index, and non-coherent modulation schemes empowered by RISs, along with other emerging RIS applications, such as next generation multiple access (NGMA), energy harvesting (EH), integrated sensing and communications (ISAC), as well as aerial and vehicular networks.

The most intriguing aspect of the RIS technology is the discussion of the technology's emerging use cases~\cite{EURASIP_RIS_all}. As we stand at the cusp of 6G, many innovative applications are in the pipeline. An RIS has the potential to revolutionize industries ranging from healthcare to transportation, fundamentally altering the way we connect, communicate, access, and process information. By shedding light on these developing applications, the article underscores the transformative role that RISs are poised to play in the future of wireless networking. Furthermore, this article covers real-world RIS prototypes and a field trial of an RIS in commercial 5G wireless networks.
\begin{figure*}[t]
\centerline{ \includegraphics[width=2\columnwidth]{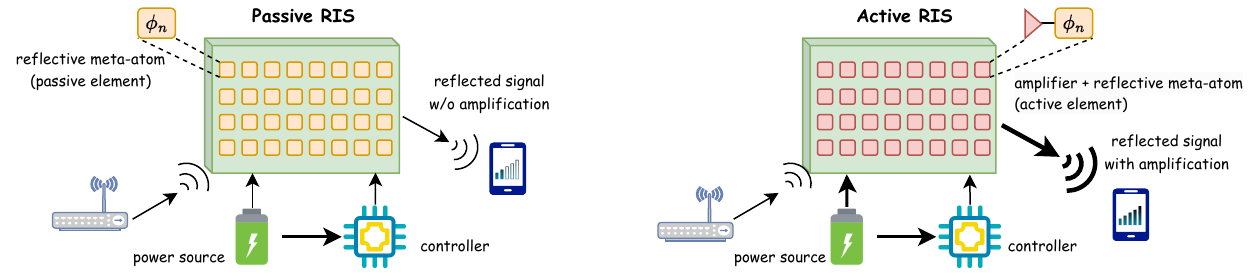}}
\caption{Comparison of a passive (left) with an active (right) RIS. The former's panel is implemented with simpler hardware but leads to lower signal strength at the receiving end(s). The effective tunable phase shift offered by each $n$-th meta-atom is represented by $\phi_n$. Links via a passive RIS suffer from multiplicative path loss, while those via an active RIS are subject to additive path loss. Both designs require a controller for the dynamic reflection configuration.
} 
\label{Active_Passive}
\end{figure*}

In summary, RISs are at the forefront of the 6G wireless revolution. This article provides a comprehensive overview of up-to-date RIS concepts, their hardware intricacies and resulting versatile operating modes, as well as the potential applications that will reshape our digital world. In this sense, this article aims to serve as a gateway to understanding how the RIS technology will redefine our connected future.

The rest of the article is organized as follows. In Section II, we provide a unified view of different RIS hardware architectures and their diverse operation modes. Next, in Section III, we cover emerging applications of RISs, followed by recent prototypes and field trials in Section IV. Finally, we discuss open problems and challenges ahead in Section V, and provide the article's conclusions in Section VI.

\section{Hardware Architectures and Operation Modes} 
The control of the propagation of EM waves has been the system design objective in many domains, ranging from medical imaging to nanolithography, and recently, wireless communications. In fact, the approaches for waves' phase and/or amplitude control via reflectarrays, metamaterials, and spatial light modulators since early $2000$s have lately inspired the RIS technology in wireless communications~\cite{Basar_2019,huang2019reconfigurable}. An RIS is a planar array of multiple ultra-thin meta-atoms (also known as unit cells or elements), each with multiple digitalized states corresponding to distinct EM responses. Tunable meta-atoms are an active area of research including diverse technologies that vary with the operating frequency, such as PIN diodes and varactors for up to millimeter wave frequencies, and liquid crystals, graphene, vanadium dioxide, memristors, and microfluidics for terahertz (THz). In the initial RIS considerations, each tunable-state meta-atom effectively contributes a phase shift on its impinging signal. To retain and/or change the state of such a reflective meta-atom requires minimal power consumption, and this is handled by an active device known as the RIS controller, which also plays the role of the RIS's interface with the wireless network. This generic almost passive and purely reflective RIS structure, known as a passive RIS, is illustrated on the left part of Fig.~\ref{Active_Passive}. Very recently, there has been increasing interest in alternative RIS hardware architectures and multi-functional capability~\cite{Tsinghua_RIS_Tutorial}, which confront with weak aspects of passive RISs, while offering additional operation capabilities.

In this section, we provide a detailed overview of the available RIS hardware architectures and their resulting modes of operation. 

\subsection{Metasurface-Based Hardware Architectures}
We next overview the latest advances in metasurface-based hardware architectures, which make use of meta-atoms of tunable EM responses as their core architectural component.

\subsubsection{RISs with Reflection Amplifiers} 
From an EM perspective, an RIS behaves as a large group of scatterers. Consequently, the end-to-end path loss of an RIS-assisted system is obtained by summing the individual path losses of the transmitter (TX)-RIS and RIS-receiver (RX) links. This is known as the double path loss effect in the RIS literature and it stems from multiplicative channels. As a result, the effectiveness of RISs reduces when they are placed far away from communicating terminals. To circumvent this, the concept of active RISs has been recently introduced~\cite{amplifying_RIS_2022,Zhang_2023}. An active RIS is a device that preserves the core benefits of a traditional passive RIS, such as operation without a transceiver radio frequency (RF) chain and lack of signal processing; however, it provides amplification to the outgoing RF signals using active reflecting components, as shown on the right part of Fig.~\ref{Active_Passive}. On the downside, the energy consumption of an active RIS is reasonably higher compared to a passive one. To this end, the RIS amplifying architecture of~\cite{amplifying_RIS_2022} deploys a single variable gain amplifier with conventional phase-tuning meta-atoms, instead of reflection amplifying meta-atoms~\cite{Zhang_2023}.  

Despite the higher power consumption, an active RIS is reported to have a higher energy efficiency in certain cases, such as when using a single power amplifier~\cite{amplifying_RIS_2022}, thanks to its significantly higher signal-to-noise ratio (SNR) and achievable data rate. It has been also shown that the multiplicative path loss effect can be transformed to an additive one thanks to the use of active RISs. After their introduction, several new designs based on active RISs have been reported in the literature. These include secure system designs, multiple access systems, wireless power transfer systems, hybrid system designs utilizing both active and passive RIS components, and so on. We expect that an active RIS might be a strong candidate to realize an efficient amplify-and-forward relay (or a network-controlled repeater) in terms of cost and complexity, and particularly, hybrid designs can be a remedy to circumvent the limited range of RIS-assisted wireless systems.

Free space optical (FSO) communication technologies are very solicited as access system technologies for the next generation of communication systems, including 5G-Advanced and 6G mobile systems, which require an increase in efficient resource utilization, and are driven by population growth. This appeal is due to the number of advantages provided by FSO technologies compared to their counterparts’ RF technologies. FSO communications utilize unlicensed bandwidths and achieve higher data transmission rates over short and long distances through the line-of-sight (LoS) path, since connection through the non-LoS link is absent. FSO systems also provide enhanced security and immunity to EM interference, low power consumption, and easy installation. However, the LoS FSO links may suffer from signal blockage due to clouds, trees, and buildings, to name only three. A solution to this dilemma is incorporating an RIS module into the FSO system to reflect the incident signal towards the targeted RX \cite{Najafi_2021}. This introduction faces a number of practical challenges, like higher attenuation due to the increase in the transmission distance and the double-fading effect that emerges from the two portions of the channel obtained after introducing the RIS module~\cite{Ndjiongue_2021}. To solve both attenuation and the double-fading effect on the transmitted signal, the research community is lately considering RIS modules with amplification capabilities. The metasurface parameters are tuned to boost the incoming signal and compensate for signal power losses caused by the introduction of the RIS module into the system. When compared to normal RIS elements, active or amplifying meta-atoms have light amplification potentials. The material used in these elements must amplify the emerging light. Liquid crystals are examples of materials that can be utilized in amplifying meta-atoms, since the manipulation of their refractive index affects both the emerging light amplitude and its orientation, thereby providing the double advantage of controlling the direction of the rising light, and more importantly, its amplitude \cite{Ndjiongue_2021}. Note that more materials could be adopted for active RISs for FSO systems. However, liquid crystals are more appealing because they are well known and a mature technology.   

\begin{figure*}[t]
	\centerline{ \includegraphics[width=2\columnwidth]{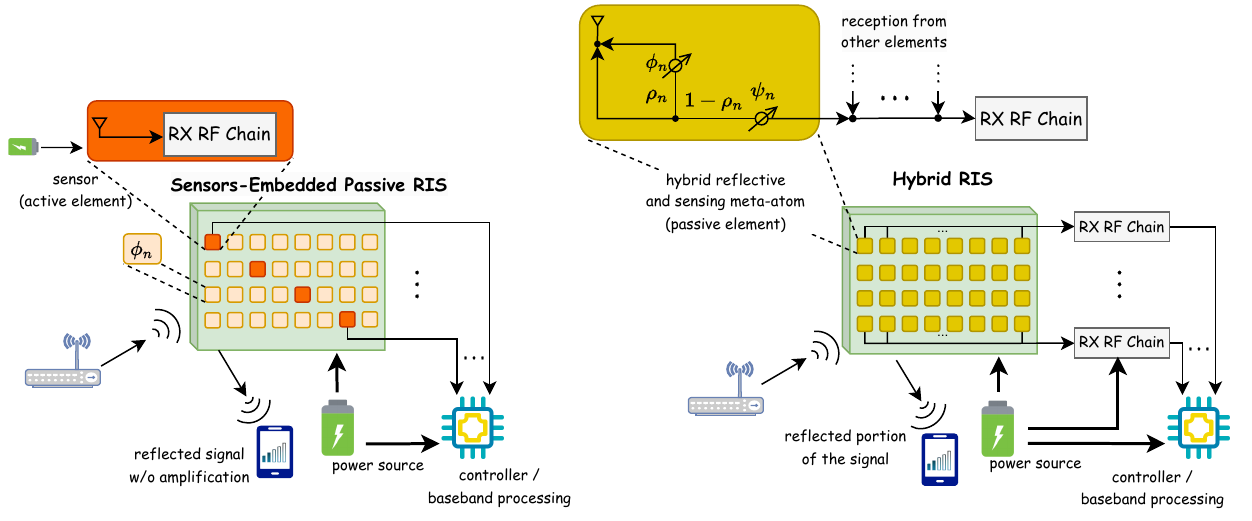}}
	\caption{ RISs equipped with both signal reception (RX) and signal reflection units. In the left hardware architecture~\cite{taha2021enabling_ALL}, the RIS panel comprises conventional reflective meta-atoms, as in passive RISs, as well as active sensing devices which enable sensing of parameters of the impinging signals that becomes available to a baseband unit, via an RX RF chain, for further processing. The right hardware architecture~\cite{alexandropoulos2021hybrid} is realized with hybrid reflective and sensing meta-atoms~\cite{alamzadeh2021reconfigurable} that split their incident signal into a portion that is reflected (after tunable phase shifting)
		in the environment, while the remainder of the signal is fed to a reception RF chain(s) for sensing and processing at the metasurface's baseband processor. $\rho_n$ represents the latter power splitting ratio at each $n$-th hybrid meta-atom, while $\psi_n$ indicates the tunable phase shift applied in the received signal. The baseband signal processing unit of both hardware architectures can be part of the RIS controller, complementing the dynamic reflection configuration management with further processing tasks (e.g., sensing and optimization).
	}  
	\label{HRIS}
\end{figure*}

\subsubsection{RIS with Signal Reception Units} 
Most commonly, to effectively operate an RIS for wireless communications, knowledge about the wireless channels between the metasurface and the communication ends is needed~\cite{Tsinghua_RIS_Tutorial}. This knowledge is, in principle, hard to acquire with a passive RIS, requiring the need for large overhead channel estimation realized at the links' end RXs, which then needs to be shared with the RIS controller for optimizing the RIS panel configuration.

To enable RISs perform estimation of parameters of their impinging signals, thus, facilitating and expediting their optimization for wireless operations, the authors in~\cite{hardware2020icassp} presented an RIS hardware architecture incorporating reception RF chains (also termed as semi-passive or receiving RIS), each usually comprising a low noise amplifier, a mixer downconverting the signal from RF to baseband, and an analog to digital converter, which are fed with the impinging signals on the RIS meta-atoms. To accomplish this functionality, those elements are each connected with sampling waveguides and tuned to a full absorption state in order to forward all the energy of their impinging signal to the corresponding waveguides, and consequently, enable the processing of the impinging signal from a baseband unit. It has been shown in~\cite{hardware2020icassp} that efficient reconstruction of parameters of an impinging signal is possible in a multi-element RIS, even with a single RF chain, using random spatial sampling and sophisticated compressed sensing techniques.

\begin{figure*}[t]
\centerline{ \includegraphics[width=2\columnwidth]{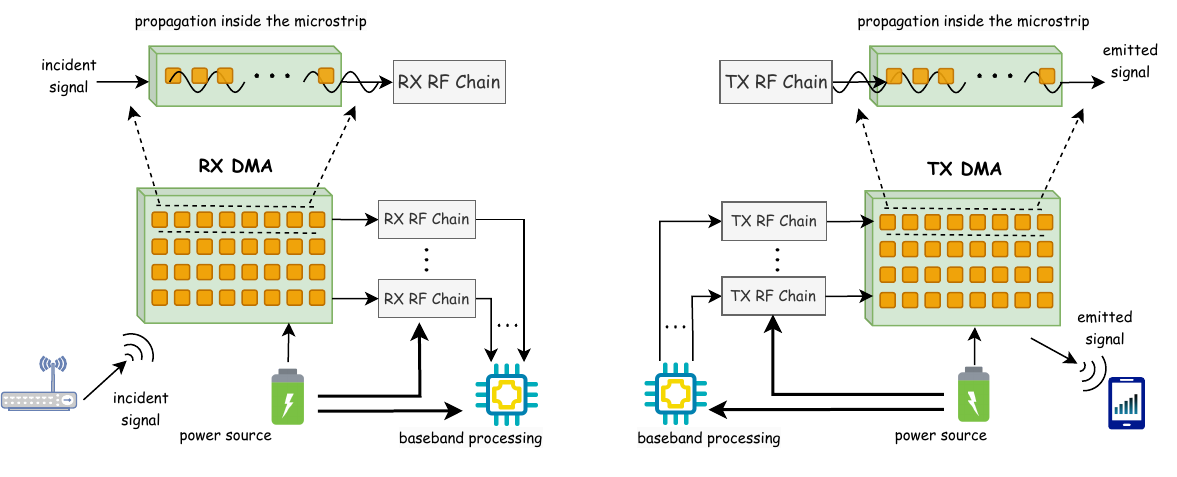}}
\caption{ Dynamic metasurface antennas (DMA) used as an extremely massive MIMO receiver (RX) and transmitter (TX). The metasurface comprises microstrips, each implemented as 1D or 2D waveguide, that include meta-atoms of tunable EM states. These elements are placed on the waveguides through which the received waveforms intended for information decoding (left) and the signals to be transmitted (right) are transferred. The TX and RX baseband processors, which respectively generate the outgoing signals and process the received signals, are connected to the waveguides through dedicated input and output ports via the TX and RX RF chains, respectively.
}  
\label{DMA_TX_RX}
\end{figure*}
\subsubsection{RISs with Signal Reception and Reflection Units} 
To exploit the aforementioned benefits of an RIS with signal reception units, the authors in \cite{taha2021enabling_ALL} proposed to replace some of the tunably reflecting meta-atoms of passive RISs with active sensing devices, as depicted in the left illustration in Fig.~\ref{HRIS}. Similar to~\cite{hardware2020icassp}, those embedded sensors will enable the RIS to estimate certain parameters of the impinging signal at its panel. The sensors' signal processing functionalities will depend on their individual computing and storage capabilities, while the overall sensing capability of this RIS architecture can be boosted using an extended RIS controller that, together with the dynamic reflection configuration management and network interfacing, performs additional signal processing tasks (e.g., sensing and optimization) via a a dedicated baseband unit. It needs to be emphasized, however, that this form of hybrid RIS does not deploy its whole panel for sensing impinging signals, but only the part where its active sensing devices are located. 

An alternative simultaneous reflecting and sensing RIS hardware architecture, termed as hybrid RIS, was recently presented in~\cite{alexandropoulos2021hybrid} and deployed for both explicit channel estimation~\cite{zhang2023channel_all} and direction-of-arrival estimation at the metasurface's side. The core component of this architecture is the hybrid reflecting and sensing meta-atom which simultaneously reflects a portion of the impinging signal, while enabling another portion of it to be sensed~\cite{alamzadeh2021reconfigurable}. This hybrid RIS architecture is illustrated on the right side of Fig.~\ref{HRIS}. This was accomplished by adding a waveguide to couple to each or to groups of meta-atoms. In particular, a via was attached to two copper traces, one to sample the signal and another to transfer the
direct current signal. A substrate integrated waveguide was used to capture the sampled wave. By changing the annular ring around the coupling via or the geometrical size of the
waveguide, the hybrid RIS can realize different coupling strengths. Consequently, each waveguide is connected to an RX RF chain, allowing the metasurface to locally process a portion of the received signals via a dedicated baseband unit, similar to the receiving RIS architecture. The elements’ coupling to waveguides implies that the incident wave is not perfectly reflected. In fact, the ratio of the reflected energy to the sensed one is determined by the coupling level. By keeping this waveguide near cutoff, its footprint can be reduced, while also reducing coupling to the sampling waveguide. It is noted that the incident wave on the hybrid RIS may couple to all sampling waveguides with different amplitudes, thus realizing a form of analog receive combining. It is finally noted that a sensors-embedded RIS can be viewed as a special case of a hybrid RIS, as one can configure some of the hybrid meta-atoms of the latter to completely absorb the impinging signals, thus, being capable to only sense. The remaining elements are set to solely reflect those signals with some desired reflection profiles. To realize a receiving RIS with a hybrid RIS, all hybrid meta-atoms need to be fixed in a fully absorbing configuration. 

\subsubsection{Dynamic Metasurface Antennas (DMA)} 
An additional emerging utilization of metasurfaces is as planar, compact,  low cost, and dynamically tunable extremely massive multiple-input multiple-output (MIMO) antenna arrays~\cite{shlezinger2020dynamic}. A DMA-based transceiver will consist of a multitude of radiating meta-atoms that can transmit and receive communication signals over the wireless channel. By dynamically tuning the EM properties of those elements, one can control the analog beampattern for transmission and reception. 

A DMA consists of a multitude of reconfigurable meta-atoms that can be used both as transmit and receive antennas, as shown in Fig.~\ref{DMA_TX_RX}. Those elements are placed in groups on one-dimensional (1D) or two-dimensional (2D) waveguides through which the signals to be transmitted and the received waveforms intended for information decoding are transferred. Such waveguides can accommodate large numbers of radiating elements, which are commonly sub-wavelength spaced, allowing each input/output port to feed a multitude of possibly coupled radiators. When a DMA is deployed as an RX, the signals captured at each meta-atom propagate through the corresponding waveguide to the output port, where they are acquired and forwarded via an RX RF chain to the baseband processing unit. In a DMA-based TX, the signals to be radiated from its meta-atoms are fed to each waveguide's input port via a TX RF chain. The relationships among the radiating signals and those captured/fed at the input/output port of each waveguide are determined by the following
two properties arising from the DMA architecture. Each meta-atom acts as a resonator whose parameters (oscillator strength, damping factor, and resonance frequency) can be dynamically configured. In each waveguide, the signal has to travel between the feed port and each meta-atom. Consequently, the signals propagating along the waveguide accumulate different frequency-dependent phases for each element. It is finally noted that DMA-based transceivers implement a form of hybrid analog and digital beamforming, since part of the processing of the transmitted and received signals is carried out in the analog domain, as an inherent byproduct of the waveguide-fed metamaterial array architecture.

\begin{figure}[!t]
\centering
\includegraphics[width=\columnwidth]{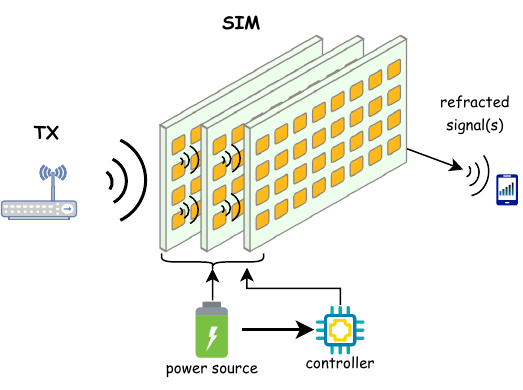}
\caption{ A stacked intelligent metasurfaces (SIM) structure comprising a 3D slab of RISs whose EM responses are managed by a dedicated controller. By appropriately designing the tunable transmissive properties of each RIS's meta-atoms, the SIM is capable to hierarchically manipulate the energy distribution of the EM waves passing through it.}
\label{SIM_structure}
\end{figure}
\subsubsection{Stacked Intelligent Metasurfaces (SIM)} By stacking an array of RISs, a three-dimensional (3D) SIM device can be obtained, as depicted in Fig.~\ref{SIM_structure}.
Following the Huygens–Fresnel principle, the EM wave passing through a meta-atom in each layer acts as a secondary point source and illuminates all the meta-atoms in the succeeding layer. Additionally, all EM waves arriving at a meta-atom in a metasurface layer are superimposed, acting as a wave incident onto this meta-atom~\cite{ICC_2023_An_Stacked}. In essence, the architecture of an SIM bears similarities to that of an artificial neural network, wherein each electronically tunable meta-atom acts as a reprogrammable artificial neuron. By appropriately designing the complex-valued transmission coefficient of these meta-atoms with the aid of a dedicated controller, the SIM gains the ability to hierarchically manipulate the energy distribution of the EM waves passing through it. Consequently, an SIM can be designed to execute various signal processing and computation tasks (e.g., image classification) in the EM wave domain. Notably, the forward propagation within the SIM occurs at the speed of light. 

\begin{figure*}[!t]
\centering
\includegraphics[width=16cm]{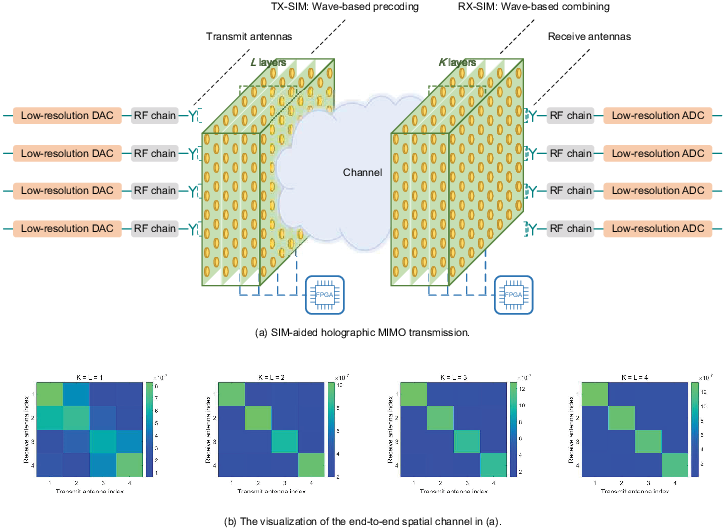}
\caption{ (a) A holographic MIMO wireless communication system comprising SIMs placed very close to a multi-antenna TX and RX, which respectively consist of $L$ and $K$ transmissive RISs. The SIMs implement analog signal processing entirely in the EM propagation domain. (b) Strength of the end-to-end wireless channel matrix for different numbers of the metasurface layers $K$ and $L$. It is demonstrated that, as the number of layers increases, the channel matrix becomes closer to diagonal.}
\label{SIM}
\end{figure*}

The SIM architecture was recently deployed in~\cite{JSAC_2022_An_Stacked} to realize a SIM-based holographic MIMO transceiver. As shown in Fig.~\ref{SIM}(a), a pair of SIMs are employed very close to the TX ($L$ transmissive RISs) and RX ($K$ transmissive RISs) antennas to implement hybrid analog and digital precoding and combining, respectively, with lower resolution digital-to-analog converters (DACs) at the TX and lower resolution analog-to-digital converters (ADCs) at the RX. In both communication sides, the analog signal processing is solely performed in the EM domain. This implies that a SIM-based holographic MIMO transceiver no longer requires computationally demanding baseband signal processing involving matrix inversion and decomposition, thus, substantially reducing the processing delay, hardware cost, as well as overall energy consumption in comparison to conventional MIMO transceiver designs~\cite{JSAC_2022_An_Stacked}. In Fig.~\ref{SIM}(b), the end-to-end spatial channel that spans from the TX to the RX SIMs is illustrated for up to $K=L=4$ layers per terminal. It can be observed that, upon increasing the numbers of metasurface layers $K$ and $L$ appropriately, the end-to-end channel matrix becomes closer to a diagonal matrix, which indicates that the overall system attains a stronger interference suppression capability, and thus, may form multiple parallel subchannels for spatial multiplexing in the physical space. This naturally boosts the achievable spectral efficiency.

\subsubsection{Simultaneously Transmitting and Reflecting (STAR)-RIS}
\begin{figure*}[!t]
  \centering
  \includegraphics[width=18.1cm]{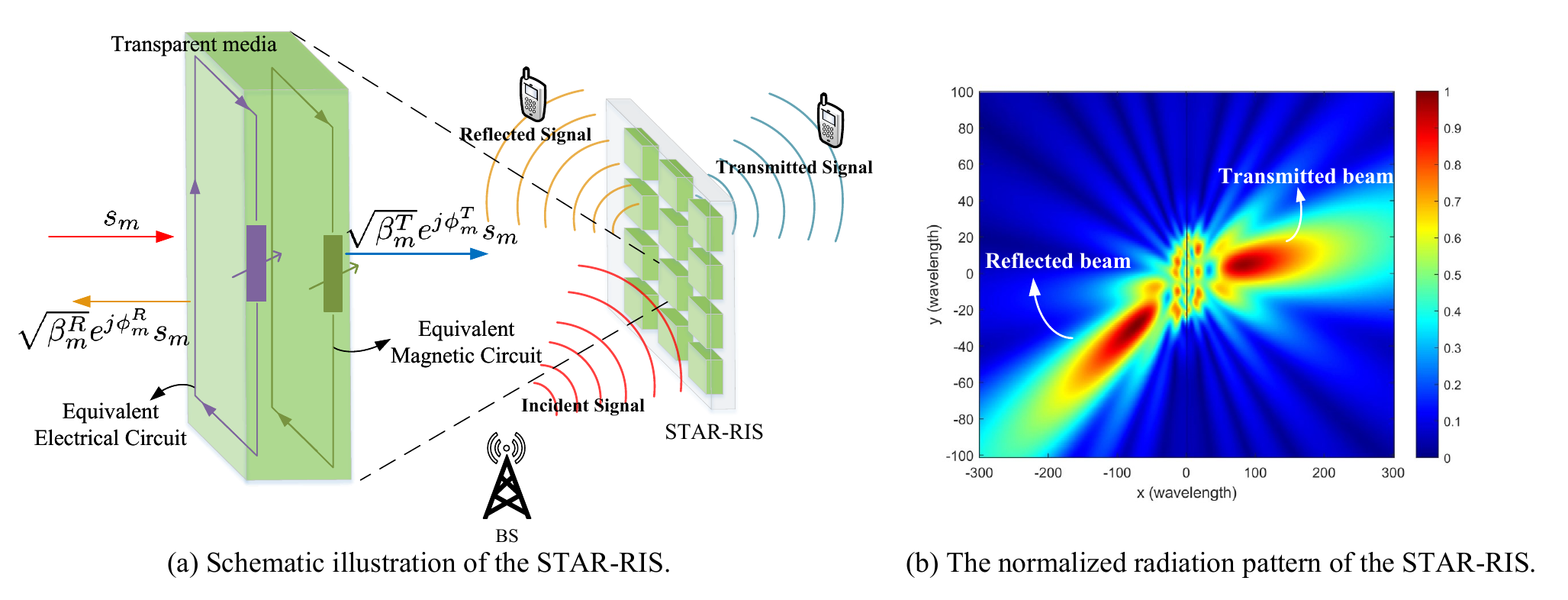}
  \caption{ Simultaneously transmitting and reflecting (STAR)-RIS: (a) the schematic diagram of a STAR meta-atom, and (b) an example radiation pattern.}
  \label{fig:STAR-RIS}
\end{figure*}
Instead of performing only reflection or refraction, a new application of using metasurface is to simultaneously achieve transmission (refraction) and reflection. Specifically, the conventional RIS configuration necessitates the co-location of the TX and RX on one side of the metasurface, thereby confining the achievable coverage to only half of the space. As a remedy, a novel approach called simultaneously transmitting and reflecting (STAR)-RIS has been introduced as a potential solution~\cite{ liu2021star}. In contrast to conventional passive RISs, STAR-RISs offer the capability to simultaneously transmit (i.e., refract) and reflect the incident signals, as depicted in Fig.~\ref{fig:STAR-RIS}(a). As a result, the requirement for the TX and RX to be situated on the same side of the metasurface is eliminated, thereby enabling a $360^{\circ}$ full-space coverage. This unique characteristic significantly enhances the design flexibility and opens new possibilities. For instance, STAR-RISs can be deployed on walls to establish seamless connectivity between two physically separated spaces, especially in millimeter wave and THz communications. Furthermore, they can also be installed on windows, enabling effective indoor--to-outdoor, and vice versa, communications. 

According to the equivalent principle in EM theory, each element of a STAR-RIS has equivalent electric and magnetic impedances. As shown in Fig.~\ref{fig:STAR-RIS}(a), two PIN or varactor diodes are required for adjusting the equivalent electrical and magnetic circuits. For a fully passive-lossless STAR-RIS, the impedances for both the electrical and magnetic circuits are purely imaginary. The two imaginary values determine the two phase shifts that are imposed on the transmitted and reflected signals, respectively. For a lossy or active STAR-RIS, the elements have non-zero real parts for their electric and magnetic impedances. By carefully tuning the amplitude and phase shift of each element, the desired beams can be generated for both transmission and reflection sides, as shown in Fig.~\ref{fig:STAR-RIS}(b) \cite{Yuanwei_2023}. However, the amplitude and phase shift control of STAR-RISs is subject to different constraints compared to conventional RISs. Some of the most prevalent constraints encountered in STAR-RISs are given as follows:
\begin{equation*}
  \beta_m^{T} + \beta_m^R = 1, \quad \cos(\phi_m^T - \phi_m^R) = 0,
\end{equation*}
where $\beta_m^T, \beta_m^R \in [0,1]$ and $\phi_m^T, \phi_m^R \in [0, 2\pi)$ are the amplitude and phase shifts achieved by the $m$-th meta-atom for the transmitted and reflected signals, respectively. Specifically, the first constraint stems from the law of energy conservation, while the second constraint is the coupled phase-shift constraint imposed by specific hardware implementations, such as passive lossless PIN diode-based implementations. Notably, an effective approach to addressing the challenging coupled phase-shift constraint has been proposed in \cite{wang2022coupled}, wherein a general optimization framework with provable optimality was proposed. It is worth mentioning that independent phase shifts of STAR-RISs can be achieved through the use of lossy or active elements, such as phased array antennas, but with high hardware cost. Metasurfaces empowered by EM metamaterials or graphene provide also a popular candidate solution for implementing STAR-RISs. Compared to the aforementioned PIN diode-based and antenna-based implementations, metasurface-based STAR-RISs can potentially offer several unique advantages, such as transparency to visible light, compatibility with high-frequency communications, and the ability to separate combined signals based on frequencies and polarizations.

\subsection{Operation Modes}
We next discuss the operational capabilities offered by the metasurface-based hardware architectures of the previous subsection that go beyond those of passive RISs (e.g., wave steering, polarizing, absorbing, filtering, and collimation).  

\subsubsection{Simultaneous Reflection and Sensing} 
This dual functionality of sensors-embedded passive RISs and hybrid RISs enables the estimation of parameters of the impinging signal at the metasurface side. In fact, the signal processing capabilities of such a metasurface constitute a design decision that depends on cost, power, and size constraints. As far as the hybrid RIS is concerned, it may be deployed to estimate the individual channels in multi-user system setups. For example, as presented in~\cite{zhang2023channel_all} for the uplink direction, the metasurface can estimate the channels between itself and multiple users via its sensed pilot observations, and then, forward this estimate to the base station over an out-of-band unidirectional control link (wired or wireless), while changing its phase configuration between the orthogonal pilot symbols’ transmissions. Alternatively, in time-duplexing division systems with orthogonal pilots used by multiple users and a base station, by configuring all hybrid meta-atoms in a full absorption (i.e., receiving) mode~\cite{hardware2020icassp}, the hybrid RIS can estimate the composite channel between itself and the multiple users as well as that with the base station, which can be again used for its reflection configuration optimization. Additionally, the hybrid RIS may be deployed for performing the estimation of the direction of arrival of impinging signals or their higher-order statistics~\cite{alexandropoulos2021hybrid}. Similar to full channel state information (CSI) estimation, this parametric estimation at the hybrid RIS side facilitates its control signaling and self optimization~\cite{Moustakas2023_RIS}, thus, contributing towards its efficient network incorporation.

\subsubsection{$360^{\circ}$ Signal Coverage} 

To facilitate wireless systems in different scenarios, STAR-RISs can work in different modes, including energy splitting (ES), mode switching (MS), and time switching (TS), which are detailed as follows:
\begin{itemize}
  \item \textbf{ES Mode}: In this mode, all STAR-RIS meta-atoms are exploited for transmission and reflection at the same time. Therefore, the energy of the incident signal is divided into two components: one for the transmitted signal and the other for the reflected signal.
  \item \textbf{MS Mode}: In this mode, each meta-atom is dedicated to either transmission or reflection. Such a binary-selection protocol makes MS mode easier to implement than the ES one, but at the cost of reduced flexibility.
  \item \textbf{TS Mode}: The transmission and reflection are not carried out simultaneously in this mode of operation. Instead, the STAR-RIS alternates between transmission and reflection in different time slots. In this case, the transmission and reflection coefficients can be independently designed. However, the TS mode necessitates precise synchronization, leading to increased complexity for its implementation. 
\end{itemize}

\begin{figure}[!t]
\centerline{ \includegraphics[width=3.65in, height=1.55in]{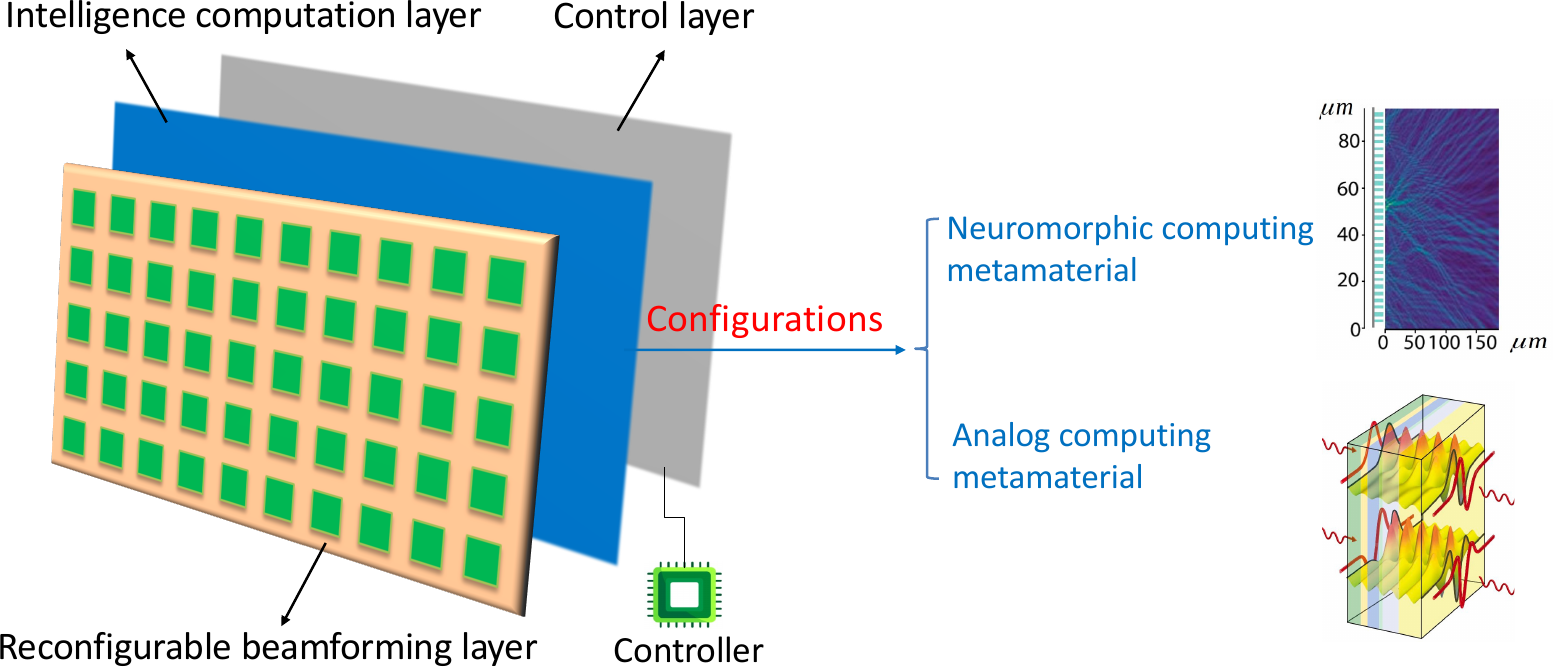}}
\caption{The architectural components of reconfigurable intelligent computational surface (RICS) comprising three layers: a reconfigurable beamforming layer, an intelligence computation layer, and a control layer.
}
\label{RICS_structure}
\end{figure}
\subsubsection{Simultaneous Reflection and Computing} 
Targeting an RIS with both computing and beamforming functionalities, the reconfigurable intelligent computational surface (RICS) concept was proposed in~\cite{RICS_WCM}. As shown in Fig.~\ref{RICS_structure}, the proposed structure is composed of: one \textit{reconfigurable beamforming layer} that is responsible for tunable signal reflection, absorption, and refraction; one \textit{intelligence computation layer} that concentrates on metamaterials-based task-oriented computing; and one \textit{control layer} that is connected with a controller responsible for the RICS parameters' configuration. To meet the various computational tasks, the intelligence computation layer of an RICS can be configured by different kind of metamaterials, e.g., neuromorphic~\cite{Neuromorphic01} or analog~\cite{Science} computing metamaterials.
\begin{figure*}[!t]
	\centering
	\includegraphics[width=2\columnwidth]{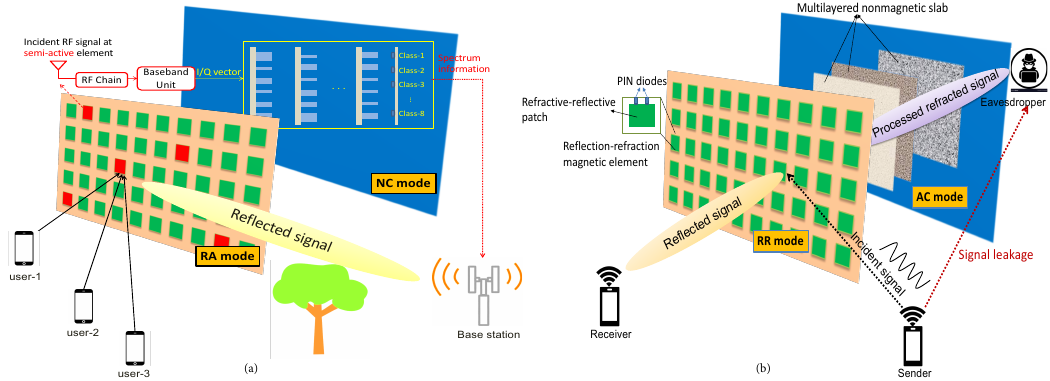}
	\caption{\small Two application designs with an RICS: (a) wireless spectrum learning, and (b) secure data transmissions.}
	\label{RICSDesign}
\end{figure*}

There exist two possible designs for RICSs~\cite{RICS_WCM}:
\begin{itemize}
\item  \textbf{Design A}: The reconfigurable beamforming and intelligence computation layers are configured as ``reflection-absorption (RA) mode" and ``neuromorphic-computing (NC) mode," respectively. In particular, in the RA mode, two types of meta-atoms, namely the passive reflecting meta-atoms and semi-active elements attached to few RX RF chains for incident signal processing similar to a receiving RIS, constitute the reconfigurable beamforming layer. In the NC mode,  the intelligence computation layer can be composed of an array of nanoribbons, which scatters the light in a way that is equivalent to artificial neural computing, as illustrated in Fig.~\ref{RICSDesign}(a). Taking wireless uplink transmission with three mobile users as an example, due to the uniqueness of the wireless signal, this computation task can be considered as a classification problem that can be addressed via this RICS design. It can be observed from the figure that, with the inferred spectrum information given by RICS, the base station can improve spectrum efficiency via allocating the wireless resources intelligently. 

\item \textbf{Design-B}: In this design, the reconfigurable beamforming and intelligence computation layers are configured as ``reflection-refraction (RR) mode" and ``analog-computing (AC) mode," respectively. Specifically, when the incident signal impinges at the reconfigurable beamforming layer, the energy can be divided into two parts, where some energy is used for reflection while the remaining energy is used for refracting the signal. By processing the refracted signal via the intelligence computation layer with AC mode, specific mathematical operations can be performed based on the incident signal. As illustrated in Fig.~\ref{RICSDesign}(b), when a sender transmits data to its RX, there exists an eavesdropper nearby trying to crack the data. With this RICS design, we can observe that an intended interfering signal can be appropriately generated by performing a mathematical operation to the refracted signal, e.g., frequency shifting, to worsen the leaked signal at the eavesdropper located at the opposite side of the RICS. Therefore, the RICS enables the exchange of confidential messages over a wireless medium in the presence of unknown eavesdroppers.
\end{itemize}
\begin{figure}[!t]
\centerline{ \includegraphics[width=1\columnwidth]{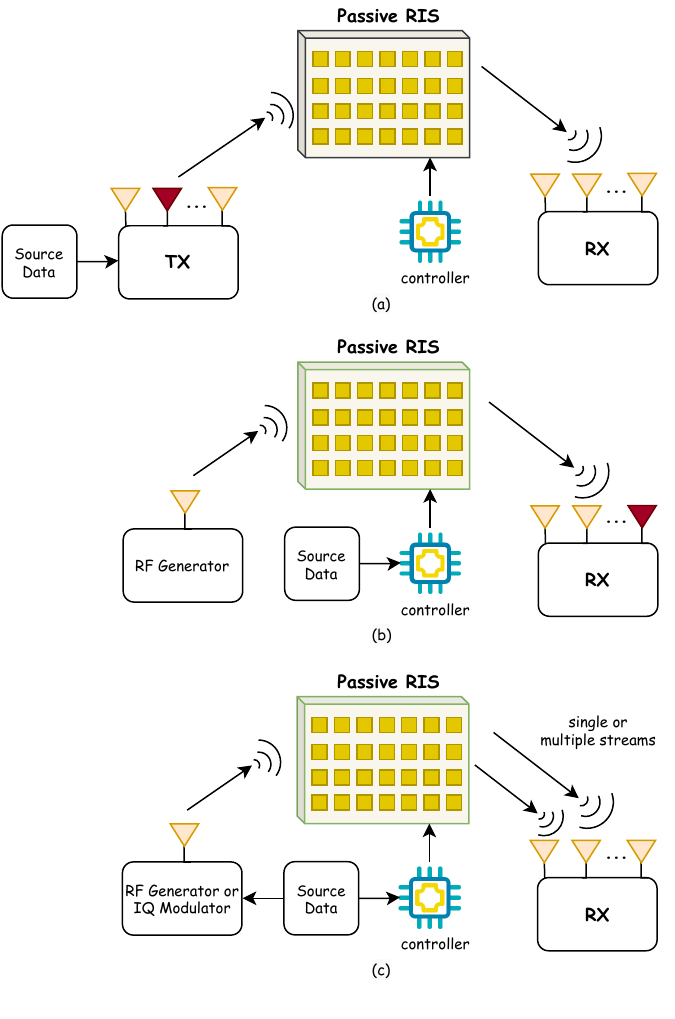}}
\caption{Conceptual index modulation (IM) and reflection modulation designs empowered by RISs. (a) transmit IM, (b) receive IM, and (c) reflection modulation.}
\label{fig:IM}
\label{IM_Reflection}
\end{figure}


\section{RIS-Inspired Wireless Applications} 
In this section, we capitalize on the different RIS hardware architectures presented in the previous section and discuss their emerging use cases and applications.

\subsection{Index Modulation (IM) and Reflection Modulation} 
The use of metasurfaces can create many opportunities for the design of next generation modulation formats thanks to their flexible architecture and ease of operation. In this context, we put forward two candidates: RIS-empowered index modulation (IM) schemes and reflection modulation realized through RISs. 

IM is a promising digital transmission method in which the indices of the available transmit entities are indexed to transmit binary information \cite{Basar_2017}. Notable four examples of the IM family are spatial modulation, orthogonal frequency division multiplexing (OFDM) with index modulation, media-based modulation, and code index modulation, which are realized by indexing transmit antennas, subcarriers, reconfigurable antenna patterns, and spreading codes, respectively. The main motivation of IM schemes is to embed information into these transmit entities to further improve either the spectral efficiency or the energy efficiency of target systems. In this context, RISs provide a new dimension for IM to further boost its attractive advantages. Specifically, as shown in Fig. \ref{fig:IM}, spatial domain IM, which is by far the most popular IM variant, can be applied at different terminals of the communication network. When applied at the TX side, as in Fig. \ref{fig:IM}(a), the task of the RIS would be boosting the overall system performance. On the other hand, using specifically designed RIS interaction matrices, receive IM can be realized to convey information by activating certain receive antennas, as shown in Fig. \ref{fig:IM}(b). In both cases, the RIS requires the knowledge of the wireless channels, while, in the latter, the RIS needs access to the information source to perform index selection at the RX. Advanced spatial IM formats, such as generalized spatial modulation and quadrature spatial modulation, can be applied at both terminals. It might be also possible to perform IM over the RIS regions by turning on and off certain groups of RIS elements, which also paves the way for reflection modulation designs. An RIS-indexed multiple access transmission scheme that utilizes dynamic phase tuning to embed multi-user information over a single carrier was recently presented in~\cite{Index_MA_RIS}. 

Reflection modulation, which is also known as reflecting modulation, RIS-based modulation, or metasurface-based modulation in the literature, is built on the idea of remodulating an unmodulated or modulated incident signal by carefully adjusting the reflection coefficients of the RIS elements, as shown in Fig. \ref{fig:IM}(c). Specifically, an RIS that is illuminated by an unmodulated RF carrier, can be utilized to create a virtual phase shift keying constellation at the RX side, enabling the implementation of a very simple signal transmission architecture without even using RF chains. Furthermore, RF-chain-free MIMO TXs are reported in the literature by dividing the RIS into multiple parts to mimic MIMO designs to convey multiple data streams to the RX. Specifically, virtual space–time coding systems utilizing Alamouti's coding, as well as experimental RIS-based MIMO architectures by using single and dual-polarized RIS designs, have been reported in the past couple of years. Alternatively, an RIS can manipulate a modulated signal as well, potentially relying on a backhaul link between the TX and the RIS, by realizing reflection modulation to send information bits to the RX. In this case, the RIS can either embed its own bits to the incident signal (separately mapped reflection modulation) or coordinate with the TX to adjust its reflection patterns according to the transmitted bits (jointly mapped reflection modulation).
\begin{figure*}
  \includegraphics[scale=0.4]{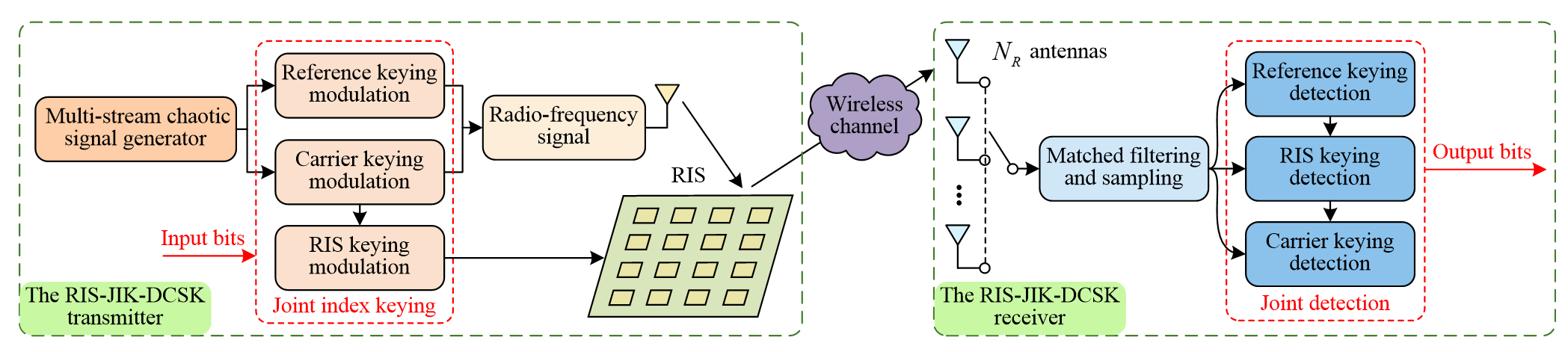}
  \caption{ Block diagram of a non-coherent $M$-ary differential chaos shift keying (RIS-JIK-MDCSK) system.}
  \label{cai}
\end{figure*}

In \cite{RX_Quad_RIS_mod}, the entire RIS was virtually partitioned into two halves to create signals with only in-phase (I) and quadrature (Q) components, respectively, and each half forms a beam to a receive antenna whose index carries the bit information. A low-complexity and non-coherent detector for this scheme was designed that measures the maximum power and polarities of the I- and Q-components of the received signals. In \cite{Space_shift_keying_RIS}, two RIS-based space shift keying (SSK) schemes were presented to improve the error and throughput performance of conventional SSK systems, respectively. The first scheme employed an RIS for beamforming targeting the maximization of the minimum squared Euclidean distance between any two decision points. The second scheme considered an RIS for Alamouti space-time block coding, enabling the RIS to transmit its own Alamouti-coded information, while reflecting the incident SSK signals to the destination node. An RIS-enhanced MIMO system with reflection pattern modulation was proposed in~\cite{Reflection_pattern_mod_RIS}. The design objective was the maximization of the average received signal power, which was accomplished through the joint optimization of the TX and RIS beamforming and phase configurations, respectively, under the assumption that the RIS's state information is statistically known at the TX.

Modern standards, such as Wi-Fi and 5G, utilize orthogonal frequency division multiplexing (OFDM) for supporting wideband communications. This indicates that future RIS-empowered modulation schemes need to support OFDM. While RIS-based reflection modulation can be particularly suitable for reduced complexity and low power applications, such as the Internet-of-Things (IoT), it has been recently reported in the literature that a frequency-modulating RIS can manipulate a single-tone carrier to create a virtual OFDM signal spanning multiple subcarriers. In addition, similar to backscatter communication systems, an RIS can embed a low-rate data stream on top of an impinging OFDM signal to convey RIS-induced information (e.g., a sensor measurement available at the RIS controller) to a nearby receiver.

\subsection{Non-Coherent Modulation} 
In RIS-aided coherent communications, channel state information (CSI) is often assumed to be perfectly estimated at the RX. However, perfect CSI cannot be available in practice. Furthermore, the complex and ever-changing channel makes the acquisition of accurate CSI unaffordable, because of the nearly passive design constraints of RISs and the large overhead in transmitting pilot signals \cite{Tsinghua_RIS_Tutorial,Cai_TWC2023}. In this context, non-coherent communications can be utilized to avoid the need for transmitting pilot signals and conducting channel estimation~\cite{Cai_CL2023}. This approach has been widely applied in various scenarios, such as secure communications \cite{Cai_TCOM2023}, underwater acoustic communications, and device-to-device communications. There are two distinct approaches for non-coherent communications. The first approach utilizes the correlation detection between consecutive received signals to recover information bits. The second approach leverages energy detection to avoid the need for pilot training overhead and simplifies the RX design, but may result in system performance degradation.

In \cite{Cai_TWC2023}, a non-coherent RIS-aided joint index keying $M$-ary differential chaos shift keying (RIS-JIK-MDCSK) system was proposed to prevent the excessive system overhead caused by channel estimation in coherent RIS-aided communication systems. In the RIS-JIK-MDCSK system, the states of the reference signal, RIS elements, and information-bearing subcarriers were jointly optimized to devise a joint index keying mechanism. Figure~\ref{cai} depicts the block diagram of the non-coherent RIS-JIK-MDCSK system. As shown in this figure, the joint index keying mechanism of the RIS-JIK-MDCSK system consists of reference keying modulation, carrier keying modulation, and RIS keying modulation. This mechanism is capable of implicitly transmitting additional information bits through the indices of these keying states, therefore, enhancing the throughput and spectral efficiency. The RIS-JIK-MDCSK system was demonstrated through simulations to have superior throughput, spectral efficiency, and error performance compared to benchmark systems, at the cost of an increased system complexity. 

In \cite{Cai_TCOM2023}, the authors proposed a non-coherent RIS-aided chaotic secure communication system that utilizes block interleaving operations to eliminate signal similarity, effectively enhancing the communication security. To recover information bits at the RX, two efficient signal detection algorithms were proposed: the sequential detection algorithm and the joint detection algorithm. The former exploits energy detection to separately recover information bits with lower complexity, while the latter utilizes the correlation detection to jointly recover those bits with a lower bit error probability. The RIS-aided chaotic secure communication system deployed an RIS in the proximity of the TX to improve the transmission quality of signals in the main channel between the TX antenna and the intended legitimate RX antenna. However, the eavesdropper channel between the TX antenna and the eavesdropping RX remains independent of the main channel, and the signal received by the eavesdropping RX cannot be improved. The analytical and simulation results presented in \cite{Cai_TCOM2023} demonstrated that the proposed RIS-aided chaotic secure communication system exhibits superior security performance compared to benchmark systems, even in scenarios where an eavesdropper attempts to wiretap the block interleaving patterns.

A zero overhead beam training scheme for RISs was proposed in~\cite{NC_RIS_1}, which relies on data transmission and reception based on non-coherent demodulation, thus, avoiding the transmission of pilot signals for channel estimation. At the RX side, the received differential data were also used for the determination of the best reflection phase profile for the RIS. It was shown in~\cite{NC_RIS_1}, by means of extensive computer simulations, that, for high mobility scenarios, non-coherent modulation is still more suitable to transmit information than classical coherent modulation. In~\cite{NC_RIS_2}, the authors focused on the uplink between a single-antenna user and a multi-antenna base station, and presented an RIS-empowered OFDM communication system based on differential phase shift keying combined with random phase configurations at the RIS. This setup avoids channel estimation and any complex RIS optimization process. It was demonstrated that the proposed RIS-assisted non-coherent modulation scheme outperforms coherent demodulation in different mobility and spatial correlation scenarios.

\subsection{Next Generation Multiple Access (NGMA)} 
The current wireless network is witnessing an exponential surge in wireless devices, particularly in Internet-of-Things (IoT) and machine-type communications (MTC) environments. Therefore, given the limited spectrum resources, it becomes crucial to develop advanced NGMA techniques that can support high data rates and accommodate massive connectivity. Due to the capability of adjusting communication channels and the favorable EM and hardware properties, RISs can provide additional DoFs, and thus, facilitate the multiple access (MA) designs from the following two aspects.
\begin{itemize}
 \item \textbf{Enhancing existing MA schemes}: The condition of communication channels plays a pivotal role in numerous popular MA schemes, such as space-division multiple access (SDMA), rate-splitting multiple access (RSMA), and non-orthogonal multiple access (NOMA). Generally speaking, different MA techniques may prefer different channel conditions and their performance heavily depends on them. Regrettably, conventional wireless networks are characterized by uncontrollable communication channels, thereby, significantly limiting the design flexibility of MA schemes. Nevertheless, the deployment of RISs constitutes a promising approach to overcome this limitation by enabling the manipulation of the over-the-air signal propagation, thus, paving the way for enhancing the MA performance. In NOMA systems, for instance, the decoder relies on the successive interference cancellation (SIC) technique to mitigate inter-user interference within the same user group. To facilitate the SIC process and guarantee system performance, careful design of the SIC decoding order and user grouping is imperative, which is subject to the channel conditions. Furthermore, in the context of multi-antenna NOMA systems, quasi-degraded communication channels serve as the desired channels for achieving the same performance as optimal dirty paper coding. Leveraging the potential of RISs, it becomes feasible to adjust the SIC decoding order based on the QoS requirement of the users, and accordingly, transform a majority of non-quasi-degraded communication channels into quasi-degraded ones \cite{liu2022reconfigurable}, thus, facilitating optimal NOMA performance. More advanced and flexible RIS-aided NOMA design can be found in \cite{liu2022reconfigurable,Cao_JSAC2023} and in the references therein.
  
  \item \textbf{Enabling new MA schemes}: When deployed as an active transceiver, i.e., when replacing a conventional antenna array with a reconfigurable metasurface, RISs provide additional DoFs that can facilitate the development of new MA schemes. This is attributed to their capability to enable signal manipulation directly in the EM domain. In contrast to conventional antenna elements, the metamaterials-based RIS elements present distinct advantages, including more compact construction and the capability to overcome the limitations associated with half-wavelength element spacing. These unique characteristics allow for the dense deployment of RIS elements, approximating a continuous radiation planar aperture, thereby, facilitating holographic MIMO~\cite{JSAC_2022_An_Stacked}. In contrast to the discrete radiation antennas, the (quasi-)continuous radiation surface can support (quasi-)continuous signal patterns, and thus, can improve orthogonality at the EM level. Recently, a holographic-pattern division multiple access (HDMA) scheme was proposed by leveraging an RIS as a transceiver \cite{deng2022hdma}. In this scheme, multiple superimposed holographic patterns were generated at the surface to convey data streams to different users. Moreover, as mentioned earlier, the utilization of an extremely large RIS can lead to the emergence of the near-field effect, which becomes more significant when the surface approaches to be continuous. In contrast to the far-field, the near-field effect introduces an additional dimension of channels in terms of distance \cite{liu2023near}, thereby, presenting a new distance domain that can be leveraged in the design of multiple access systems.
\end{itemize}

In summary, one or multiple RISs can facilitate the design of NGMA schemes by operating either as passive relays between transceivers to reshape the MA channels or as active transceivers themselves offering additional DoFs. Moreover, there exists the potential to integrate these two roles of RISs, i.e., the functions of passive relaying and active transmission/reception, to further advance NGMA designs, which constitutes a promising research direction.

\subsection{Integrated Sensing and Communications (ISAC)} 
Third generation partnership project (3GPP) Release 16 introduced dedicated 5G positioning reference signals and measurements, as well as new features boosting the estimation accuracy of time- and angle-based localization. The 5G New Radio provides up to $100$ MHz in frequency range 1 (FR1) and $400$ MHz in frequency range 2 (FR2), contributing further to the localization accuracy: the delay error variance decreases in the order of the square of the increasing bandwidth. In addition, large antenna apertures offer high angular resolution. In particular, the variance of angle estimation is proportional to the inverse square of the antenna spacing. Furthermore, the number of rows and columns of an antenna array gives a cubic decrease in the angle estimate variances. Later 3GPP releases up to the upcoming Release 18 frozen state capitalize on these items to offer improved positioning of active network devices. 

ISAC constitutes a novel network service~\cite{ISAC_Mag_2021} which is expected to employ 3GPP's technical specification group on service and system aspects in early Release $19$ discussions. It envisions to deploy the same frequency bands for both communications and sensing purposes, and the multiplexing capabilities of current wireless systems (e.g., waveforms and MIMO) will efficiently trade-off communications and sensing services. RF sensing mainly deals with the super-resolution detection and tracking of passive objects in the wireless environment, e.g., gesture capturing and activity recognition, as well as immersive applications, like digital twins, that require context information from the environment to dynamically reconstruct it. In addition, RF sensing can be used to improve the performance of communication systems, e.g., more accurate beamforming, faster beam failure recovery, and less overhead when tracking the wireless channel.
\begin{figure}
  \includegraphics[scale=0.75]{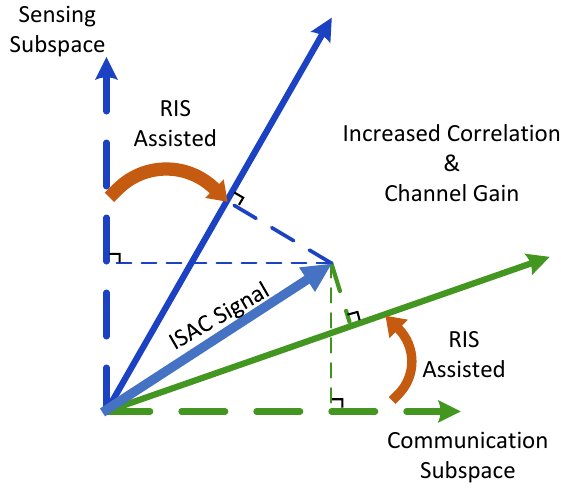}
  \caption{An RIS can be optimized to both expand and rotate the sensing and communications subspaces, thus, maximizing the collective gains from simultaneous sensing and communications operations~\cite{RIS_ISAC_SPM}.}
  \label{fig:RIS_ISAC}
\end{figure}

\begin{figure*}[htbp]
  \includegraphics[scale=1]{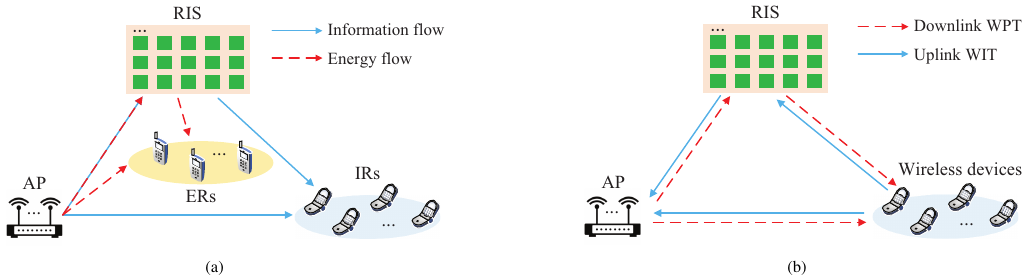}
    \caption{Illustration of the integration of an RIS into the two main application paradigms of RF-based energy harvesting: (a) An RIS-aided simultaneous wireless information and power transfer (SWIPT) system; (b) An RIS-aided wireless-powered communication network (WPCN).}
    \label{fig:model}
\end{figure*}
The capability of RISs to manipulate the propagation of EM waves and engineer virtual LoS conditions has been also recently leveraged to create additional signal propagation paths (i.e., signal path diversity) for the purposes of RF localization, sensing, and ISAC~\cite{RIS_ISAC_SPM}. RIS-aided localization capitalizes on the additional reflections offered by programmable metasurfaces to improve positioning and sensing accuracy, or even, enable these services in settings with a limited number of transmitting terminals and non-LoS environments. In fact, RISs can enable the localization of active users in several scenarios~\cite{Keykhosravi2022infeasible}, including cases where there is no access point or base station available in the system. Additional sensing-related applications of RISs that have been recently explored in the literature include environmental mapping, non-LoS sensing at millimeter waves and THz, and monostatic or bistatic radar, where it has been shown that metasurfaces contribute to the energy-efficient optimization of the illuminated power in geographical areas of interest, thus, improving therein the identification capability of passive/active targets~\cite{RIS_ISAC_SPM}. As far as ISAC is concerned, it has been recently theoretically shown that RISs can be optimized to offer two jointly significant features~\cite{RIS_ISAC_SPM}. Similar to all their other considerations (i.e., communications and localization individually), they can create additional channel paths, and consequently, additional channel gains, for both sensing and communications. In addition, RISs can be dynamically configured to maximize the correlation between the sensing and communications wireless channels, which consequently maximizes the gains of the simultaneous realization of these operations. Figure~\ref{fig:RIS_ISAC} considers an initial scenario of weakly coupled sensing and communication channels (e.g., when the communication user is an unmanned-aerial-vehicle (UAV), while the target to sense is a ground vehicle), and illustrates the RIS capability to jointly expand and rotate the subspaces of sensing and communication functionalities.  


\subsection{Energy Harvesting} 
Harvesting energy from ambient RF signals is envisioned as a green and cost-effective solution to extend the lifetime of energy-constrained wireless networks. However, due to impairments in wireless channels, such as path loss, RF-based EH typically suffers from low energy efficiency, especially when the transmission distance is long. Despite the potential enhancement of received power at the energy RX (ER) through increased radiation power at the TX, the end-to-end efficiency remains unimproved. Fortunately, RIS is a promising technology to mitigate this issue. Specifically, the fine-grained reflect beams, created by leveraging intelligent passive reflections over the large aperture of the RIS, can effectively compensate for the severe signal attenuation over long transmission distances. This enables the establishment of enhanced wireless charging zones for ERs in the RIS's vicinity, which holds practical significance in enlarging the coverage range of wireless power transfer (WPT). Also, for traditional WPT systems facing substantial efficiency degradation caused by blockages, the RIS can be deployed to create virtual LoS links between the TX and the ERs to bypass blockages, thereby improving the EH performance of these systems. Moreover, a preliminary RIS prototype with $1$-bit programmable reflecting elements has been recently proposed and implemented in \cite{2019_Tran_WPT}, showcasing experimentally the ability of an RIS to enhance the efficiency of practical WPT systems.  

Metasurfaces can be integrated into the two main application paradigms of RF-based EH, i.e., simultaneous wireless information and power transfer (SWIPT) as well as wireless-powered communication network (WPCN), as illustrated in Fig. \ref{fig:model}. In SWIPT systems, based on the physical feature of RF signals carrying both information and energy, wireless information transmission (WIT) and WPT can be carried out simultaneously using the same RF signals. Existing works on integrating RISs into SWIPT systems have recently gained some meaningful insights. For example, for the case of separated ERs and information RXs (IRs), it was shown in \cite{2020_Qingqing_SWIPT_QoS} that compared to the case without RIS(s), deploying metasurfaces not only reduces the transmit power required at the access point, but also effectively decreases the number of required energy beams, thereby greatly simplifying the transmit beamforming design. In addition, for the case of co-located ERs and IRs, the authors in~\cite{2023_Ying_IFC} considered three practical transmission schemes for SWIPT in a multi-user interference channel, namely the hybrid time switching and power splitting scheme, the time division multiple access (TDMA) scheme, and the TDMA-D scheme. The simulation results indicated that the introduction of RISs can rewrite the performance comparison results among these schemes and that the hybrid time switching and power splitting scheme assisted by RISs with appropriate numbers of reflecting elements performs the best in terms of achievable sum rate. On the other hand, in WPCNs, the wireless devices first harvest energy from the received RF signals and then utilize their harvested energy to transmit information. An RIS can be employed to improve the amount of energy harvested at the devices during the downlink WPT phase and enhance the devices' rate performance during the uplink WIT phase. Particularly, the authors of \cite{2021_Qingqing_NOMA} revealed that, using different RIS phase shifts for the downlink WPT and the uplink WIT yields no benefits to their considered WPCN, thus, reducing the number of optimization variables and also lowering the feedback signaling overhead. What's more, the utilization of an RIS in the WPCN offers a twofold advantage: enhancing the system throughput, while simultaneously reducing the energy consumption at the access point. 

It is worth noting that the above discussions are based on purely passive RISs. Compared to a passive RIS, an active RIS shows greater advantages in enhancing the received power at the ERs and extending the operating range of WPT, since it can alleviate the product path loss attenuation inherent in the RIS-aided cascaded channel. However, the non-negligible RIS-amplified noise power, while helpful for WPT, it is detrimental for WIT. Despite this, existing works have validated the superiority of adopting an active RIS over a passive RIS in SWIPT systems and WPCNs. Surprisingly, the numerical results in \cite{2022_Piao_Active} demonstrated that the active RIS-aided WPCN can achieve higher throughput with a lower total system energy consumption than its passive RIS counterpart. 

\begin{figure*}[!ht]
	\centering
	\includegraphics[width=0.55\textwidth]{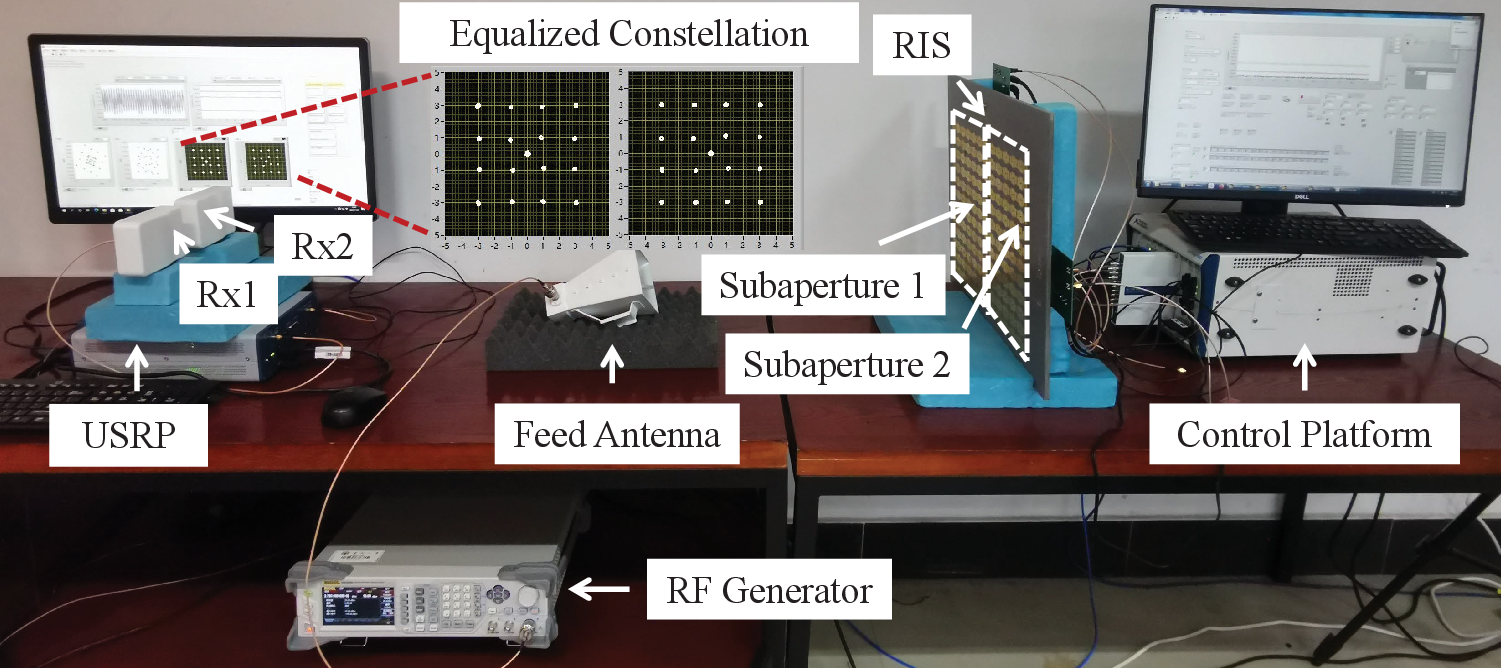}
	\caption{ The prototype of an RIS-based $ 2\times 2$ IM-MIMO-QAM wireless communication system operating at $2.7$ GHz.}
	\label{fig:testing_system}
\end{figure*}

\subsection{Aerial and Vehicular Networks} 
While dedicated short-range and cellular-based wireless communication technologies can support current vehicular applications, facilitating future advanced vehicle-to-everything services, such as vehicle platooning and autonomous driving, presents primarily two challenges: 1) how to overcome blockage in complex environments of vehicular communication scenarios; and 2) how to deal with rapidly time-varying and non-stationary channels due to high vehicle mobility \cite{6G-V2X}. Fortunately, incorporating RISs provides an effective solution due to their capability to produce additional virtual LoS channels and improve channel conditions via adjusting the phase shifts of the RIS elements. In general, current research focuses on two network architectures: deploying RISs in the environment (e.g., on the surfaces of roadside infrastructures and buildings) and mounting RIS on vehicles. Statically deployed roadside RISs can extend the coverage area for roadside base station-vehicle (RBS-vehicle) and vehicle-to-vehicle communications. For example, communication signals can be transmitted via unblocked cascaded paths to vehicle users on perpendicular roadways or shadow regions of other vehicles, thereby, enhancing the received SNR and coverage probability. In addition, on-vehicle RISs can facilitate communication for in-car equipment or nearby vehicles by intelligently reflecting/refracting incident signals to ``slow down'' multipath fading or to compensate for Doppler effects \cite{RIS-Doppler-0}. As described in \cite{RIS-slowfading}, an on-vehicle RIS facilitates communication between a remote base station and a user inside a high-speed vehicle. A two-stage transmission protocol for channel estimation and RIS phase shift optimization was proposed to improve channel quality. In addition, the simulation results demonstrated the effectiveness of mitigating channel fading fluctuations and the superiority of the proposed scheme over deploying a roadside static RIS. 

The majority of the aforementioned RIS applications in vehicular networks are based on terrestrial communications. Extending in the vertical dimension, aerial networks have two distinct advantages over terrestrial vehicular communications: 1) higher probability of providing LoS channels arising from high altitudes; and 2) 3D mobility with additional design degrees of freedom \cite{WuUAVoverview}. Similar to applications in vehicular networks, RISs can significantly improve the performance of aerial networks, especially UAV-based networks. On the one hand, by deploying RISs around terrestrial communication nodes, not only the transmitted/received signal strength of UAV can be enhanced, but the propulsion energy can also be saved. For instance, joint reflecting/refracting coefficients and placement/trajectory optimization can be designed to save the energy consumption of aerial nodes for movement, while satisfying quality-of-service requirements for IoT applications \cite{WuUAVoverview}. On the other hand, RISs mounted on aerial nodes can also contribute to signal enhancement and coverage extension as a mobile relay or an aerial user. Notably, compared to mounting an RIS in a single aerial node, multiple nodes carrying an aerial RIS deserves more research interest due to its unique potentials, such as UAV-swarm-enabled aerial RIS (SARIS) \cite{SARIS}. For example, the size of an RIS can sometimes be relatively large to support ultra-high data rate transmissions, while a single UAV cannot afford to carry such a heavy payload. Nevertheless, a large RIS can be virtually carried by a UAV swarm, where each UAV only needs to carry a relatively small portion of the large distributed RIS. Although this decomposition of a large RIS remains largely unexplored, the optimal 3D deployment of a SARIS has been preliminarily studied in \cite{SARIS}, concluding that the optimal deployments of a SARIS and that of a terrestrial RIS are distinct. 

Although the RIS technology is promising for supporting future vehicular and aerial communication services, real-time channel information acquisition is very challenging. Moreover, there are still various unresolved issues when both networks are integrated, such as the complex interference topology and real-time resource allocation.

\section{RIS Prototypes and Field Trials} 
The past couple of years have witnessed a surge of experimental studies with RISs, thanks to the vast availability of RIS prototypes worldwide. Notable examples of fabricated RISs include those from industrial and academic bodies, such as ZTE, Huawei, NEC Laboratories, Southeast University, Ruhr University Bochum, University of Surrey, University of Glasgow and others, as well as multi-partner projects on the topic, e.g., the Horizon $2020$ RISE-6G project and the Smart Networks and Services phase $1$ TERRAMETA project. In particular, different bands, including below $6$ GHz, mmWaves, and sub-THz, have been considered as well as the operation modes of reflection, sensing/reception, and transmission. It is undeniable, however, that the vast majority of available fabricated RISs is predominantly based on passive RIS architectures with internal controller hardware. In addition, the primary focus of state-of-the-art RIS experimentation has been to showcase the feasibility of coverage extension with RISs, where various investigation have reported improvements of around $10$ dB in the received signal strength with a carefully optimized RIS. Another set of experimentation dealt with RISs deployed as a part of a transmitter with the intention to explore the potential of RISs to create virtual MIMO systems and implement RF-chain-free transmitters. Very recently, there have also appeared studies focusing on codebook-based RIS configurations targeting fast reconfiguration of RIS phase profiles with respect to the angles of the end terminals.
	
	Against this exciting background, this section focuses on two emerging RIS applications, namely, RIS-based IM prototyping and RIS-empowered coverage extension for commercial 5G cellular networks.
	
\begin{figure*}[!ht]
	\centering
	\includegraphics[width=0.55\textwidth]{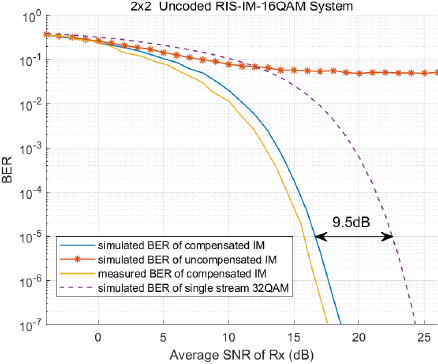}
	\caption{Measured and simulated bit error rate performance of the RIS-based $2 \times 2$ IM-QAM system.}
	\label{fig:results}
\end{figure*}

\begin{figure*}[!ht]
	\centering
	\includegraphics[scale=0.30]{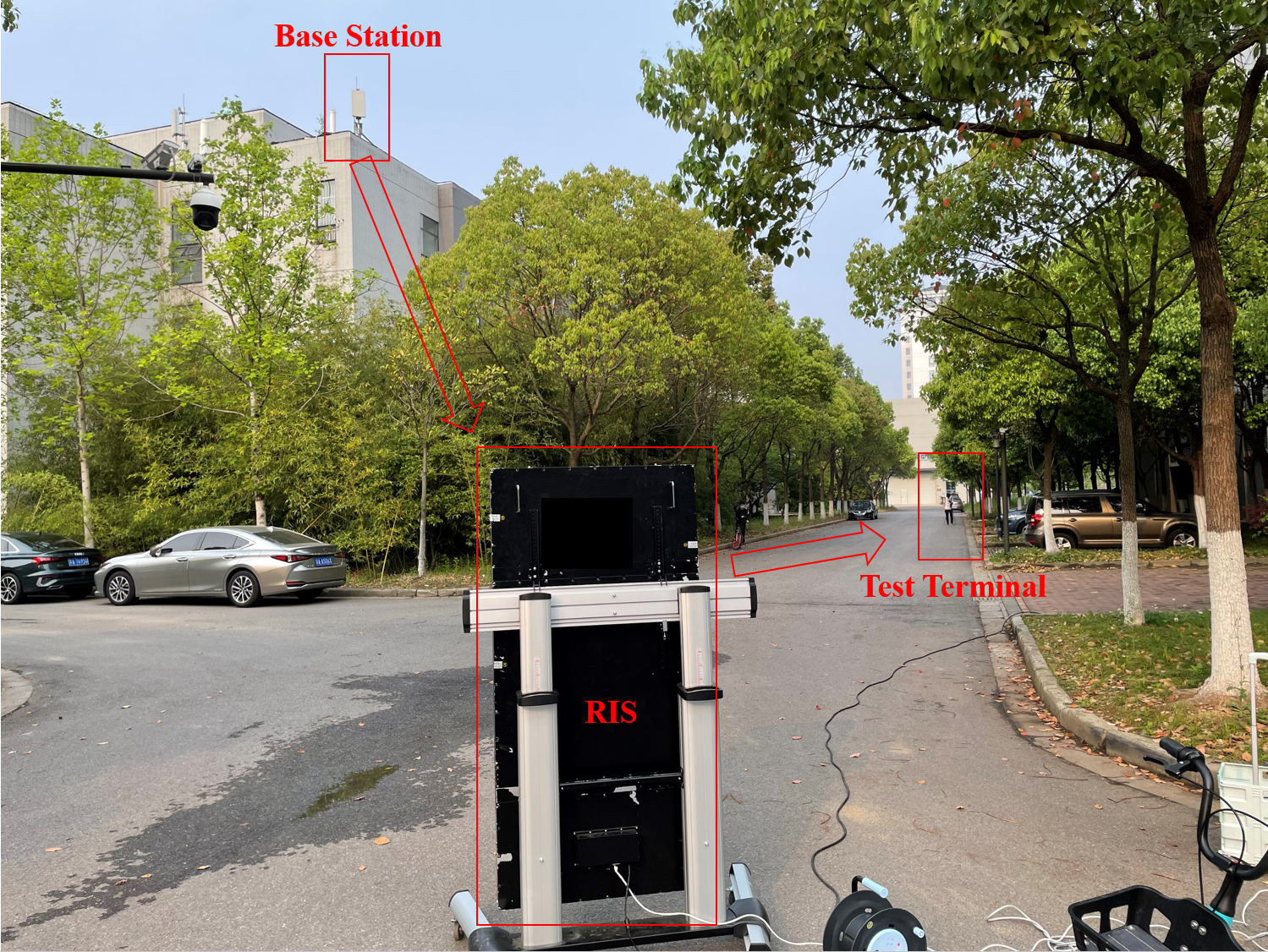}
	\caption{ The outdoor scenario of RIS-based 5G coverage extension in Jiulong Lake Campus of Southeast University in China.}
	\label{fig：RIS_trial}
\end{figure*}

\subsection{RIS-Based Index Modulation}

An RIS-based $2\times2$ IM MIMO communication prototype with quadrature amplitude modulation (QAM) was recently developed in~\cite{Liu2022} and is illustrated in Fig.~\ref{fig:testing_system}. The TX is located on the right side of the figure and consists of an RIS, a control platform, a feed antenna, and RF signal generator. The RX is located on the left side, consisting of two RX antennas, a universal software radio peripheral (USRP), and a desktop computer. The RIS is composed of $12$ unit elements, divided into left and right subsurfaces. In each symbol time slot, the single-side subsurface is activated according to the index bit, while the other subsurface is loaded with a constant control voltage to stop the QAM modulation. The DACs convert the digital control sequence mapped from the baseband digital bits into external control voltage, to adjust the amplitude and phase of the order $-1$ harmonic component of the carrier wave. RIS-based IM can be achieved using non-linear modulation and a constellation compensation method. At the RX, the two received signals are down-converted by the USRP platform, and the computer performs synchronization, least-squares channel estimation, and IM symbol detection based on maximum likelihood. The received antenna index and transmitted symbol are jointly demodulated and converted into a serial bit stream.

The conducted experimental results in~\cite{Liu2022} showcased that, compared to an RIS-aided single-antenna system with $16$-QAM and the same symbol rate, the designed RIS-based $ 2\times 2$ IM-MIMO-QAM system promotes the bit rate by $25\%$, mainly due to the implicit transmission dimension brought by the index bit. When the IM-MIMO-QAM scheme was carried out without constellation compensation, the actual generated constellation diagram differed from the ideal constellation diagram to varying degrees in the clockwise direction and the bit error rate performance deteriorated sharply. The reason for this is the significant phase deviation between the uncompensated and standard constellation diagrams. The constellation obtained according to the proposed compensation method in~\cite{Liu2022} was consistent with the standard $16$-QAM constellation diagram. Compared to a $32$-QAM single-antenna transmission system with the same bit rate, it was found that the IM-MIMO-QAM transmission system saves about $9.5$ dB of average SNR, i.e., transmit power, when the bit error rate reaches $10^{-5}$ as illustrated in Fig.~\ref{fig:results}.

\subsection{RIS-Based 5G Coverage Enhancement}
In light of the widespread adoption of the 5G technology, conventional coverage enhancement technologies are facing severe challenges due to high power consumption and hardware cost. RISs, being capable to flexibly regulate incident signals and construct intelligent communication environments with low cost, are considered as a potential efficient technology for signal coverage extension.

Conducting RIS field trials in a 5G commercial network is of paramount importance for validating the effectiveness of the technology and quantifying its performance across diverse real-world network scenarios. The authors in~\cite{RIS_5G_trial} presented a 5G commercial network coverage enhancement prototype system based on an RIS. In this system, the 5G base station of China Mobile served as the TX for the downlink, while a smartphone connected to the 5G network operated as the signal testing terminal, recording the received signal power from the selected base station. An RIS was deployed to reflect its incident signal to a set of predetermined directions, by performing reflective beamforming according to its elements' possible phase configurations. The prototype system implemented an RIS phase configuration optimization algorithm, which was based on a closed-loop feedback mechanism to enhance the overall reflective beamforming performance.

The outdoor scenario of the RIS-based 5G coverage enhancement field trial in~\cite{RIS_5G_trial} is depicted in Fig.~\ref{fig：RIS_trial}. The scenario focused on a weak coverage case where the base station was deployed at a high altitude, in particular, on the top corner of a building, and the signal test terminal was placed opposite to the base station orientation at the building's floor level. The LoS transmission path between the base station and the test terminal was blocked by a row of trees. As shown in the figure, to create a virtual LoS path, an RIS was deployed at the crossing of two roads in face of the antenna array of the base station. At that location, the path between the RIS and the test terminal was unobstructed; see the signal reflection path indicated by a red arrow in the figure. In summary, the steps for the RIS optimization in~\cite{RIS_5G_trial} were the following: First, the RIS was placed so as to enable a LoS path between itself and the base station and another LoS path between itself and the test terminal. Then, the RIS ran the phase configuration optimization algorithm. The measurement results for this field trial revealed that, when compared to the benchmark without the metasurface, the deployment of the RIS, combined with the use of appropriate phase configuration, leads to a notable increase in the received signal power, in particular, a gain of $4.03$ dB. It was also demonstrated that the proposed RIS phase configuration optimization outperforms an approach based on RIS configuration sweeping among sample phase configurations. In~\cite{2022_5G_trial}, additional field-trial investigations further confirmed that deploying RISs into current 5G networks can improve user experience, extend signal coverage, and enhance throughput in various urban scenarios. 

%

\section{Open Challenges and Future Directions}

Capitalizing on the previously described rich background on emerging RIS hardware architectures, as well as their operation modes and recent applications, this section sheds light on relevant open problems and potential future directions.

Passive RIS architectures stand out with low power consumption and low implementation costs. On the other hand, considering the rise of network-controlled (smart) repeaters, the design of novel RIS-assisted systems via active and hybrid metasurfaces to improve the RIS illumination area and consequently the overall energy efficiency of the network is an exciting research direction. In this context, the power consumption and implementation cost of their active elements need to be efficiently traded off in conjunction with their application scenarios. Designing novel mixed-type RIS architectures at this stage is a promising research area aiming to unveil a compromise between passive and wholly active RIS architectures. Exploiting RISs with reception units adds a new dimension to RIS applications by enabling RIS terminals to estimate parameters of the impinging signals. On the one hand, since this architecture requires RX RF chains, this feature might render them incapable to compete with smart repeaters in terms of achievable communication performance despite entailing a similar implementation cost. On the other hand, RISs embedding active signal reception units, much less in number compared with programmable metamaterial units, will enable more efficient optimization of their reflection operation as well as facilitate sensing functionalities.   

Despite its remarkable advantages, the practical implementation of the SIM architecture is hindered by some unique challenges, such as experimental verification and the introduction of mathematically tractable SIM models together with efficient SIM configuration methods, which are still in their infancy and deserve further investigation. Furthermore, a SIM cannot be deployed as a plain RIS in the environment and requires embedded TX and RX units to manipulate EM waves effectively. STAR-RISs also shine with their ability to provide full $360^{\circ}$ coverage; however, selecting the proper operating mode is challenging, considering different requirements of the users and the complexity of their practical implementation. At this stage, emerging use cases are being evaluated for STAR-RISs, with the dominant one being outdoor-to-indoor coverage extension, which might be vital for future systems and standards employing mmWaves or even sub-THz frequencies.

The limited data rate of RIS-based reflection modulation designs is a critical problem for their effective operation and can limit their consideration mainly in IoT-type applications. In this context, designing novel systems that can reliably embed more information bits into dedicated entities in RIS-assisted communication systems, even on active and hybrid RISs, is an exciting research direction. Furthermore, creating new OFDM-based index/reflection modulation schemes is a promising direction for integrating RIS-empowered modulation concepts with legacy wireless systems and standards.

Despite their vast potential, RIS-based EH and aerial/vehicular networks aided by RISs might be challenging to implement in practice due to the high mobility of RXs. Notably, the placement of an RIS in a given setup plays a vital role in delicate use cases such as EH systems, and effective protocols need to be developed for the reliable transmission of information and energy to intended users. Ground RISs are the first step for aerial systems since RIS-mounted flying platforms face serious challenges, such as seamless integration to these fragile platforms and high path loss due to more considerable distances. To this end, conformal RISs constitute another promising direction for coating moving aerial or terrestrial vehicles. 

In terms of practical widespread deployment of RISs, the design of true standalone smart RIS solutions that can adapt according to user locations, wireless channels, and system requirements is necessary to eliminate the need for large-overhead control signaling with the network. Self-configured hybrid RISs can contribute to this goal. In addition, the efficient placement of RISs in smart wireless environments is still an open critical problem for real-world scenarios. Furthermore, more realistic experimental setups should be considered to operate RISs without requiring dedicated horn antennas or placing them at artificial locations in the network. All in all, exciting and challenging research problems still exist to unlock the true potential of RISs for wireless communications, localization, sensing, and their integration.

\section{Conclusions}
RIS-enabled wireless communications and sensing is a game-changing family of technologies in the rapidly evolving realm of 6G wireless networks. Offering precise control over the propagation environment, the RIS paradigm is vital for implementing seamless, sustainable, and cost-coefficient wireless applications. Our exploration in this article covered RIS hardware intricacies, including various architectures and operating modes. We have also highlighted the emerging RIS applications, such as reflection modulation, NGMA, ISAC, and EH, which are poised to reshape various industries. This article aims to serve as a gateway to understanding how RISs will redefine our connected future in the 6G era. We have delved into specific design aspects of RISs and RIS-enabled smart wireless environments, providing further insights into the capabilities and potential challenges of the RIS technology.
\bibliographystyle{IEEEtran}
\bibliography{bib_2023}

\end{document}